\documentclass[10pt,twocolumn]{article}
\usepackage{ol2}
\usepackage{graphicx,epsfig}

\begin{document}

\twocolumn[ 

\title{Evanescent Airy beams}


\author{Andrey V. Novitsky$^{1*}$ and Denis V. Novitsky$^{2}$}

\address{
$^1$ Department of Theoretical Physics, Belarusian State University,
Nezavisimosti~Avenue~4, 220030 Minsk, Belarus\\
$^2$ B.I. Stepanov Institute of Physics, National Academy of
Sciences of Belarus, Nezavisimosti~Avenue~68, 220072 Minsk, Belarus}

\email{$^*$Corresponding author: andrey.novitsky@tut.by}

\begin{abstract}
In this Letter we propose the concept of the evanescent Airy beam.
We analyze the structure of the ideal evanescent Airy beam, which
initial profile has the Airy form, while the spectral decomposition
consists of only evanescent partial waves. Also, we discuss the
refraction of the Airy beam through the interface and investigate
the field of the transmitted evanescent Airy beam.
\end{abstract}


 ] 

The propagating Airy beams are well defined by the paraxial wave
equation \cite{Berry,Sivil07PRL}
\begin{equation}
\frac{\partial u}{\partial z} + \frac{1}{2 k} \frac{\partial^2
u}{\partial z^2} = 0.
\end{equation}
The above equation provides the solution in the form of a
non-diffracting Airy beam, which accelerates in $x$ direction (see
Refs.
\cite{Sivil07PRL,Sivil07,Besieris,Sivil08,SiviloglouOPN,Broky08}).
For highly non-paraxial beams both non-diffractivity and
acceleration are not the case \cite{Novitsky}. Such beams possess
comparable inputs of propagating and evanescent waves. The
evanescent Airy beams are studied in this Letter.

At first we will define the field of the evanescent Airy beam. We
start with the solution of Maxwell's equations for the TE-polarized
plane wave
\begin{eqnarray}
E_y (x,z,k_x) = c(k_x) {\rm e}^{i k_x x + i k_z z},
\end{eqnarray}
where $E_y$ is the electric field orthogonal to the plane of
wavevectors ($x$, $z$), $c(k_x)$ is an amplitude, $k_x$ and $k_z =
\sqrt{k_0^2 \varepsilon \mu - k_x^2}$ are the transverse and
longitudinal wavenumbers (their names are defined with respect to
the propagation direction of the Airy beam), $k_0 = \omega / c$ is
the wavenumber in vacuum, $\varepsilon$ and $\mu$ are the dielectric
permittivity and magnetic permeability of the medium, respectively.
The appropriate choice of the amplitude $c(k_x)$ yields the needed
initial intensity distribution (beam's profile). If
\begin{eqnarray}
c (q) = \frac{1}{2 \pi} A {\rm e}^{-a q^2} {\rm e}^{i(q^3 - 3 a^2 q
- i a^3)/3}, \label{amplc}
\end{eqnarray}
the initial beam is the Airy beam ${\rm Ai}(x/x_0) \exp(a x/x_0)$,
where $a$ is a decay factor limiting the beam energy at $x<0$, $x_0$
is an arbitrary transverse scale \cite{Sivil07,Besieris}. Here and
in what below the dimensionless parameters are used: transverse
wavenumber $q = k_x x_0$, wavenumber in vacuum $\chi = k_0 x_0$,
spatial ranges $\tilde{x} = x/x_0$ and $\tilde{z} = z/x_0$. Fourier
transform results in the non-paraxial wave solution
\begin{eqnarray}
E_y (x,z) = \int_{-\infty}^\infty  c(q) {\rm e}^{i q \tilde{x} + i
\sqrt{\chi^2 \varepsilon \mu - q^2} \tilde{z}} dq. \label{Airi_gen}
\end{eqnarray}

The spectrum of transverse wavenumbers can be divided into two
domains. For $-n \chi<q<n \chi$ the partial plane waves are the
propagating waves ($n = \sqrt{\varepsilon \mu}$ is the refractive
index). In the rest $q$-region the waves are evanescent, because
they are characterized by the imaginary longitudinal wavenumber. In
paraxial approximation ($\chi$ is great) the spectrum of propagating
waves is wide and these waves dominate the evanescent ones. In the
opposite case, when $\chi<<1$, the spectrum of propagating waves is
narrow and we can neglect these waves. Finally, the solution for
evanescent waves can be written with disregarding propagating waves
as
\begin{eqnarray}
E_y (x,z) = \frac{A}{2 \pi} \int_{-\infty}^\infty  {\rm e}^{i q
\tilde{x} - |q| \tilde{z}} {\rm e}^{-a q^2} {\rm e}^{i(q^3 - 3 a^2 q
- i a^3)/3} dq. \label{Airi_evan}
\end{eqnarray}

The above integral can be interpreted as the {\it evanescent Airy
beam} because of two reasons. First, each of the partial plane waves
in the Fourier integral (\ref{Airi_evan}) is evanescent. Second, the
initial (at $\tilde{z} = 0$) beam has the ideal limiting Airy
profile ${\rm Ai}(\tilde{x}) \exp(a \tilde{x})$. If compared with an
evanescent Bessel beam \cite{NovitskyBessel}, the situation appears
to be very similar. Bessel beam can be also presented as the
superposition of the plane waves, the wavevectors forming a cone.
That is why the longitudinal wavenumber of all the partial plane
waves is the same. In the case of the evanescent Bessel beam this
wavenumber becomes a complex number. Since it is equal for each
partial wave, it can be carried out the integral. For an Airy beam
the longitudinal wavenumbers of the partial plane waves are
different and we need to integrate over transverse wavenumber $q$.

\begin{figure}[t!]
\includegraphics[scale=0.5, clip=]{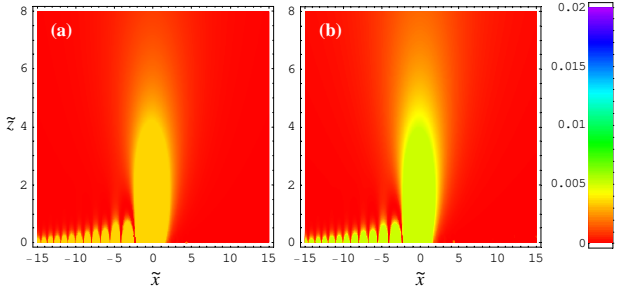}
\caption{Comparison of (a) an ideal evanescent beam
(\ref{Airi_evan}) and (b) realistic one (\ref{Airi_gen}). Intensity
$|E_y/A_1|^2$ is calculated. The decay parameter $a=0.05$.
Parameters for figure (b): $n=1$, $\chi=0.1$. }\label{fig:1}
\end{figure}

Eq. (\ref{Airi_evan}) describes the ideal evanescent Airy beam. In
real situations, the beam contains also the propagating waves in the
narrow $q$-domain, exactly $-n \chi<q<n \chi$. One more
approximation is the replacing of the expression $\sqrt{\chi^2
\varepsilon \mu - q^2}$ by ${\rm i} |q|$, because we assume small
parameter $\chi$. This leads to the independence of the electric
field (\ref{Airi_evan}) on the refractive index of the medium. In
Fig.~\ref{fig:1} we compare the Airy beams calculated using Eq.
(\ref{Airi_evan}) and using the exact formula (\ref{Airi_gen}). One
observes that the lobes of the beam are formed by the evanescent
waves. The propagating waves added in Fig \ref{fig:1}(b) only
strengthen the beam, but do not influence on its structure. The
intensity of propagating waves is distributed homogeneously over the
transverse coordinate $\tilde{x}$. Therefore, all the lobes are
intensified approximately equally. The main lobe propagates farther
than that for purely evanescent waves due to the propagating waves.
If parameter $\chi$ diminishes, the contribution of the propagating
waves becomes smaller and both figures (a) and (b) become more and
more similar. Condition $\chi<<1$ implies large wavelength compared
with the transverse scale, i.e. $\lambda >> 2 \pi x_0$. With such
condition the evanescent profile shown in Fig. \ref{fig:1} can be
experimentally realized.

\begin{figure}
\includegraphics[scale=0.5, clip=]{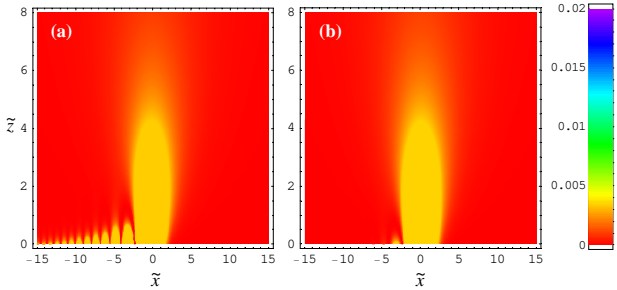}
\caption{Evanescent Airy beams at decay parameters (a) $a=0.1$ and
(b) $a=0.5$. }\label{fig:2}
\end{figure}

Evanescent Airy beams have the only parameter in Eq.
(\ref{Airi_evan}), it is the decay parameter $a$. From Fig.
\ref{fig:2} we conclude that small decays provide the side lobes of
the evanescent Airy beam. Until $a = 0.1$ the evanescent beams are
very similar, what can be seen from comparison of Fig.
\ref{fig:1}(a) and Fig. \ref{fig:2}(a). The beam structure stays
approximately the same. This can be analyzed using the spectrum
(\ref{Airi_evan}). At small $a$ the main input takes the cubic term
${\rm e}^{i q^3/3}$, which is responsible for the fringes. However,
great decay parameters $a$ significantly change the beam structure.
The side lobes almost disappear at $a=0.5$ (see Fig.
\ref{fig:2}(b)). It should be noted that the main lobe is
approximately the same as in Fig. \ref{fig:2}(a), i.e. its
properties are not determined by the parameter $a$. The side lobes
disappear in the full agreement with the results of Ref.
\cite{Besieris}, the authors of which showed that the limited Airy
beam at large $a$ lose the fringes. Addressing again to Eq.
(\ref{Airi_evan}) we conclude that at great $a$ the cubic term does
not play an important part in comparison with the other terms. This
assertion is valid for both propagating and evanescent Airy beams.

To illustrate the concept of the evanescent Airy beam we consider
transmission of the incident propagating Airy beam through the
interface between two non-magnetic media, the first medium with
refractive index $n$, the second one with refractive index 1. Taking
only propagating plane waves in the spectrum of the incident beam we
describe it as follows
\begin{eqnarray}
E_y (x,z) = \int_{-n\chi}^{n\chi}  c(q) {\rm e}^{i q \tilde{x} + i
\sqrt{\chi^2 n^2 - q^2} \tilde{z}} dq. \label{Airi_inc}
\end{eqnarray}
Each plane wave in this spectrum transmits through the interface
$\tilde{z} = \tilde{z}_0$ according to the Fresnel formula
\begin{eqnarray}
\tau (q) = \frac{2 \zeta_1(q)}{\zeta_1(q) + \zeta_2(q)} {\rm e}^{i
(\zeta_1(q) - \zeta_2(q)) \tilde{z}_0}, \label{transmit}
\end{eqnarray}
where $\zeta_1 = \sqrt{\chi^2 n^2 - q^2}$ and $\zeta_2 =
\sqrt{\chi^2 - q^2}$ are the normalized longitudinal wavenumbers.

After transmission, some of the waves (with $-\chi<q<\chi$) stay
propagating waves. The others are subjected to the total internal
reflection, therefore, they become evanescent waves. So, the
electric field of the transmitted Airy beam can be written as
\begin{eqnarray}
E_y (x,z) = \int_{-n\chi}^{n\chi} \tau (q) c(q) {\rm e}^{i q
\tilde{x} + i \sqrt{\chi^2 - q^2} \tilde{z}} dq. \label{Airi_trans}
\end{eqnarray}

\begin{figure}[t!]
\includegraphics[scale=0.5, clip=]{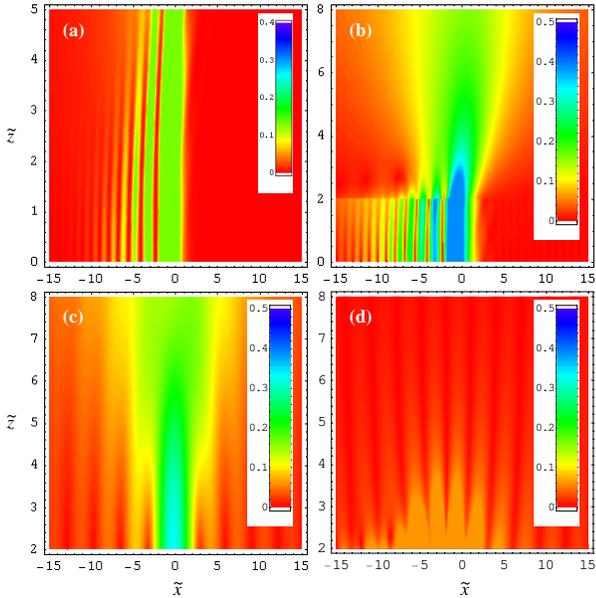}
\caption{ Transmission of the Airy beam through the interface
between dielectric media with $n=3.5$ ($0<\tilde{z}<2$) and $n=1$
($\tilde{z}>2$). (a) Intensity $I = n |E_y/A_1|^2$ of the incident
Airy beam. (b) Transmission of the incident Airy beam through the
interface. (c) Transmitted propagating waves. (d) Transmitted
evanescent waves. For the figures (b)--(d), the quantity $\sqrt{I}$
is plotted. Parameters: $a=0.1$, $\chi=1$.}\label{fig:3}
\end{figure}

The numerical simulation of the refraction of the Airy beam is
demonstrated in Fig. \ref{fig:3}. The ``free'' beam, which is not
undergone to the interaction, has some bright side lobes (see Fig.
\ref{fig:3}(a)). We place the interface at $\tilde{z}_0=2$ and
observe the transmitted Airy beam. Reflected beam is not the case
for our investigation.

Both incident and transmitted beams are shown in Fig.
\ref{fig:3}(b). We have calculated the amplitude (the square root of
intensity) in this figure, because the amplitude of the beam profile
is more vivid than intensity. The side lobes of the transmitted Airy
beam rapidly decay. The main lobe spreads, so that it swallows up
the nearest side lobes. Transmitted beam loses its accelerating
property and shifts a bit to positive coordinates $\tilde{x}$ after
refraction.

So, what are the reasons of losing accelerating properties? At
first, it is confined spectrum of the incident beam, which takes
into account only propagating waves. Therefore, transmitted beam has
no all evanescent components, and the intensity pattern of the
evanescent waves (and of the whole beam) is distorted. Second, we
consider great refractive index for the incident medium, so that
much plane waves of the spectrum undergo the total internal
reflection and become evanescent. Acceleration is the property of
the propagating waves (in usual paraxial approximation all partial
plane waves are propagating). Evanescent waves, as it has been
discussed before, does not provide the acceleration of the beam.
Third, the side lobes, which are essential for the accelerating Airy
beams, are very weak, therefore, the beam spreads and is not more
similar to the Airy beam.

\begin{figure}[t!]
\includegraphics[scale=0.5, clip=]{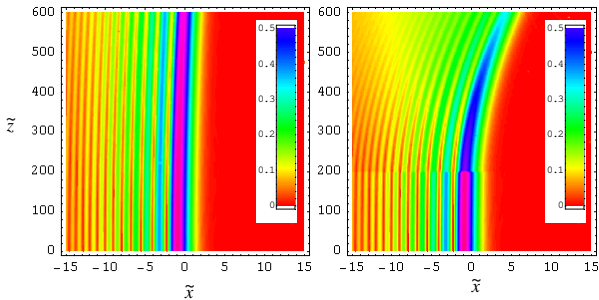}
\caption{ (a) Propagation of the Airy beam without refraction. (b)
Transmission of the Airy beam through the interface between
dielectric media with $n=3.5$ ($0<\tilde{z}<200$) and $n=1$
($\tilde{z}>200$). Parameters: $a=0.1$, $\chi=100$. }\label{fig:4}
\end{figure}

The spreading of the propagating waves (Fig. \ref{fig:3}(c)) is why
the whole transmitted beam spreads. The side lobes of propagating
waves are very weak. Except the side lobes on the left from the main
lobe, the symmetric lobes due to the confined spectrum of the beam
appear on the right. The picture of the intensity distribution
becomes almost symmetric. The propagating waves carry the energy
from the interface. As well-known, it is not the case for the
evanescent waves. The intensity pattern of the evanescent waves has
the fan-like form (see Fig. \ref{fig:3} (d)). It is mostly symmetric
and correspond to the rapidly decaying beam. However, the pattern
does not look like in Fig. \ref{fig:2} (a). It can be understood
because of the following reasons. First, the beam has large
parameter $\chi$, which does not allow us to exclude the propagating
waves from the plane wave spectrum. Second, the spectrum of the
evanescent waves is very restricted by the non-evanescent properties
of the incident beam. Third, the interaction of the incident beam
with the interface changes its ideal form. These items are
responsible for the fan-like form of the transmitted evanescent
beam.

If refractive index $n$ increases, the transmission provides
qualitatively the same results as those in Fig. \ref{fig:3}.
However, the side lobes of the total transmitted and evanescent
beams become more pronounced. If the incident Airy beam is paraxial
and formed with both propagating and evanescent waves, it keeps the
Airy structure after refraction as shown in Fig. \ref{fig:4}. It
should be noted that the transmitted beam accelerates faster than
the incident one, because it propagates in the air.

In conclusion, we have demonstrated the ideal evanescent Airy beam,
which consists of only evanescent partial plane waves and has the
Airy profile at the initial plane $z=0$. We have discussed the
limitations of such a beam and concluded that it can be realized
only in specific situation (it should be small parameter $\chi$ and
all the wavenumbers $q$, from $-\infty$ to $\infty$, should be taken
into account). We have considered the refraction of the incident
propagating Airy beam through the interface and analyzed the
transmitted evanescent beam and its deviation from the ideal one. We
believe that the obtained theoretical results can be useful for the
aims of microscopy.

\end{document}